\documentclass[proof]{WileyASNA-v1}

\articletype{Original Paper}
\received{1 July 2020}

\usepackage{amsmath}
\usepackage{amsfonts}
\usepackage{amssymb}
\usepackage{graphicx}
\usepackage{epsfig}
\usepackage{mathrsfs}
\usepackage{bm}
\usepackage{microtype}
\usepackage{hyperref}
\usepackage{mathrsfs}
\usepackage{sirimacro}

\usepackage{xcolor}

\pdfoutput=1

\begin{document}

\title{Extreme Primordial Black Holes}
\author[1,2]{Siri Chongchitnan}  
\author[3]{Teeraparb Chantavat}
\author[4]{Jenna Zunder}

\authormark{CHONGCHITNAN \textsc{et al}}


\address[1]{\orgdiv{Mathematics Institute}, \orgname{University of Warwick}, \orgaddress{Zeeman Building, CV4 7AL, \state{Coventry}, \country{United Kingdom}}}

\address[2]{\orgdiv{E. A. Milne Centre for Astrophysics}, \orgname{University of Hull}, \orgaddress{Cottingham Rd., HU6 7RX \state{Hull}, \country{United Kingdom}}}

\address[3]{\orgdiv{ Institute for Fundamental Study}, \orgname{Naresuan University}, \orgaddress{65000 \state{Phitsanulok}, \country{Thailand}}}

\address[4]{\orgdiv{Department of Mathematics}, \orgname{University of York}, \orgaddress{Heslington, YO10 5DD\state{York}, \country{United Kingdom}}}

\corres{*Siri Chongchitnan, Mathematics Institute, University of Warwick. \email{siri.chongchitnan@warwick.ac.uk}}

\abstract[Abstract]{
We present a formalism for calculating the probability distribution of the most massive primordial black holes (PBHs) expected within an observational volume. We show how current observational upper bounds on the fraction of PBHs in dark matter translate to constraints on extreme masses of primordial black holes. We demonstrate the power of our formalism via a case study, and argue that our formalism can be used to produce extreme-value distributions for a wide range of PBH formation theories.
}


\maketitle

\section{Introduction}

Primordial black holes (PBHs) originate from large inflationary perturbations that subsequently collapse into black holes in the early Universe (for reviews, see \cite{sasaki,garcia,carr,kashlinsky}). LIGO gravitational wave events\footnote{\url{https://www.ligo.org/detections.php}} over the past few years have given rise to the resurgence of PBHs not only as a viable dark matter candidate, but also as potential pregenitors of massive black holes ($\gtrsim30M_\odot$) that can typically  give rise to the observed amplitude of gravitational waves.

Given an inflationary scenario, it would be useful to predict the mass of the most massive PBHs expected within a given observational volume. Such a calculation would serve as an additional observational test of competing inflationary theories. The primary aim of this work is to present such a framework, whilst demonstrating the method for a particular model of PBH formation.

The framework discussed is based on previous work by one of us \citep{me_void1,me_void2} in the context of extreme cosmic voids, as well as previous work by  \cite{harrison,harrison2} on extreme galaxy clusters. Our main result will be the probability density function (pdf) for the most massive PBHs expected in an observational volume. We will apply the framework to a simple model of PBH formation and demonstrate the soundness of the calculations.

Throughout this work we will use the cosmological parameters for the LCDM model from \ii{Planck} \citep{planck18}.


\section{Fraction of the Universe in PBHs}

In this section we will derive an expression for the cosmological abundance of PBHs, namely
\ba \Omega\sub{PBH}={\rho\sub{PBH}\over\rho\sub{crit}},\ea
where $\rho\sub{PBH}$ is the mean cosmic density in PBHs, and $\rho\sub{crit}$ is the critical density. Typically we will be interested in the abundance of PBHs within a certain mass range (say, $\Omega\sub{pbh}(>M)$, \ie\ the fraction of the Universe in PBHs of mass greater than $M$). The PBH abundance naturally depends on how primordial perturbations were generated (\eg the shape of the primordial power spectrum of curvature perturbations), details of the PBH collapse mechanism (\eg  structure formation theory), and thermodynamical conditions during the radiation era when PBHs were formed. We will obtain an expression for $\Omega\sub{PBH}$ that depends on all these factors.

One viable approach to begin modelling the PBH abundance is to use the Press-Schechter (PS) theory \citep{press}, but with a modification of the collapse threshold. This approach has been widely used in previous work to model PBH abundances (\eg \cite{chong_pbh, ballesteros, wang, byrnes, young}). We present the key equations below. In section \ref{peakt}, we present a different approach based on Peak Theory.

In the PS formalism, the probability that a region within a window function of size $R$, containing mass $M$, has density contrast in the range $[\delta, \delta+d\delta]$ is given by the Gaussian distribution
\ba P(\delta)d\delta={1\over\sqrt{2\pi}}{1\over\sigma}e^{-\delta^2/2\sigma^2}\D\delta,\lab{ps}\ea
where $\sigma$ is the variance of the primordial density perturbations $\delta$ smoothed on scale $R$. Assuming that PBHs originate from Fourier modes that re-entered the Hubble radius shortly after inflation ends (\ie\ during radiation era, when $R$ becomes comparable to $k^{-1}=(aH)^{-1}$), $\sigma$ can be expressed as  \cite{liddle}
\ba \sigma^2(k)&=\int_{-\infty}^\infty W^2(qk^{-1})\mc{P}_\delta(q) \D\ln q\\
&=\int_{-\infty}^{\infty} {16\over81}\,W^2(qk^{-1})(qk^{-1})^4 T^2(q, k^{-1}) \mc{P}_{\mc{R}}(q)\D \ln q.\lab{1681}\ea
In the above equations,  $\mc{P}_{\delta}$ and $\mc{P}_{\mc{R}}$ are, respectively, the primordial density and curvature power spectra;  $W$ is the Fourier-space window function chosen to be Gaussian\footnote{See \cite{ando} for an interesting study of how window functions affect the inferred PBH abundances.} ($W(x)=e^{-x^2/2}$);  $T$ is the transfer function given by:
\ba T(q, \tau) =  {3\over y^3}\bkt{\sin y-y\cos y}, \quad y \equiv {q\tau\over \sqrt3}.\ea


Numerical simulations suggest that the initial mass, $M$, of a PBH formed when density perturbation of wavenumber $k$ re-enters the Hubble radius, is known to be a fraction of the total mass, $M_H$, within the Hubble volume ($M_H$ is usually called the `horizon mass'). In this work, we follow \cite{musco}  in modelling $M$ as
\ba M= K(\delta-\delta_c)^{\gamma} M_H.\lab{fraction}\ea
where we take $\delta_c=0.45$ (the threshold overdensity for collapse into a PBH during radiation era), with $K=3.3$ and $\gamma=0.36$. (See section \ref{profile} for further discussion  of the values of $\delta_c$ and $K$.)



For a given Hubble volume with horizon mass $M_H$, the corresponding temperature, $T$, satisfies the equation 
\ba M_H = 12\bkt{m\sub{Pl}\over\sqrt{8\pi}}^{3}\bkt{10\over g_{*,\rho}(T)}^{1/2}T^{-2},\ea 
\citep{wang} where $m\sub{Pl}$ is the Planck mass, and the effective degree of freedom $g_{*,\rho}(T)$, corresponding to energy density $\rho$, can be numerically obtained as described in \cite{saikawa}. The latter reference also gave the fitting function for the effective degree of freedom $g_{*,s}(T)$ corresponding to entropy $s$, which we will also need. 


Using the extended PS formalism, one obtains the following expression for $\beta_{M_H}$, the fraction of PBHs  within a Hubble volume containing  mass $M_H$ \citep{niemeyer, byrnes}
\ba\beta_{M_H} &= 2\int_{\delta_c}^\infty {M\over M_H} P(\delta)\D\delta\notag\\
&=\int_{-\infty}^{\infty} B_{M_H}(M) \D\ln M.\ea
The factor of 2 is the usual Press-Schechter correction stemming from the possibility of PBHs formed through a cloud-in-cloud collapse \citep{bcek}. The integrand $B_{M_H}(M)$ can be interpreted as the probability density function (pdf) for PBH masses on logarithmic scale at formation time. Using \re{ps} and \re{fraction}, one finds
\ba B_{M_H}(M)&= {K\over \sqrt{2\pi}\gamma\sigma(k_H)}\mu^{1+1/\gamma}\exp\bkt{-{1\over 2\sigma^2(k_H)}\bkt{\delta_c+\mu^{1/\gamma}}^2},\\
\mu&\equiv {M\over KM_H},\lab{muu}\\
{k_H\over \text{Mpc}^{-1}}&=3.745\times10^{6}\bkt{M_H\over M_\odot}^{-1/2}\bkts{g_{*,\rho}(T(M_H))\over106.75}^{1/4}\ldots\nn\\
&\ff\times\bkts{g_{*,s}(T(M_H))\over106.75}^{-1/3}.\lab{kh}
\ea

We next consider an important quantity  $f(M)$, the present-day fraction of dark matter in the form of PBHs of mass $M$. For our purposes, the expression for $f(M)$ can be expressed as \cite{byrnes2}:
\ba f(M) &\equiv {1\over\Omega\sub{CDM}}\diff{\Omega\sub{PBH}}{\log M}\notag\\
&={\Omega_m\over\Omega\sub{CDM}}\int_{-\infty}^\infty 
\tau(M_H) B_{M_H}(M) \D \log M_H,\lab{fM}\\
\tau(M_H)&\equiv {g_{*,\rho}(T(M_H))\over g_{*,\rho}(T\sub{eq})}{g_{*,s}(T\sub{eq})\over g_{*,s}(T(M_H))} {T(M_H)\over T\sub{eq}},\lab{thermo}\ea
where $\Omega\sub{CDM}$  and  $\Omega_m$ are the cosmic density parameters for cold dark matter and total matter (CDM+baryons) respectively.  The first integral \re{fM} is an integration over all horizon masses. Since a range of PBH masses are expected to form within a given horizon mass $M_H$, this integral picks out the fraction of those PBHs with mass $M$. The thermodynamical factor, $\tau(M_H)$, relates the formation-time variables to present-day observables. Using standard expressions for the evolution of cosmic densities during the matter and radiation era, and the fact that $\rho\sub{PBH}/\rho\sim T^{-1}$ up to matter-radiation equality, we can intuitively understand the appearance of temperature at  matter-radiation equality ($T\sub{eq}$) and at formation time ($T(M_H)$) in \re{thermo}. See \cite{inomata} for a detailed derivation.

Once we have calculated the PBH fraction, $f(M)$, the total present-day fraction of PBHs in dark matter can be calculated by integrating over all PBH masses,
\ba f\sub{PBH}=\int_{\log M\sub{min}}^\infty f(M) \D\log M,\lab{fint}\ea
(we will discuss $M\sub{min}$ in the next section). Finally, the fraction of the Universe in PBHs of mass $>M$ can simply be integrated as
\ba\Omega\sub{PBH}(>M)=\Omega\sub{CDM}\int_{\log M}^\infty f(M^\pr) \D\log M^\pr.\lab{om}\ea



\section{PBH number count}

In analogy with the abundance of massive galaxy clusters (see \eg \cite{mobook} for a pedagogical treatment), the differential number density of PBHs at present time (\ie\ the PBH `mass function') can be expressed as:
\ba \diff{n}{\log M}&= -{\bar\rho\over M}\diff{\Omega\sub{PBH}(>M)}{\log M} ={\bar\rho\over M}\Omega\sub{CDM}f(M),  \lab{massfunc}\ea
where $\bar\rho$ is the present-day mean cosmic density. In an observational volume covering the fraction $f\sub{sky}$ of the sky up to redshift $z$, we would find the total number of PBHs to be
\ba
N\sub{tot}(z)= f\sub{sky} \int^{z}_{0}(1+z^\pr)^3 \textrm{d}z^\pr\int_{\log M\sub{min}(z^\pr)}^\infty  \textrm{d}\log M  \ff \diff{V}{z^\pr}   \diff{n}{\log M},\lab{Nto}
\ea
where d$V/$d$z$ is the Hubble volume element given by
\ba\diff{V}{z}&= {4\pi\over H(z)}\bkt{\int_0^z{\D z^\pr\over H(z^\pr)}}^2, \\
H(z)&\approx H_0\bkts{\Omega_m(1+z)^3+\Omega_r(1+z)^4+\Omega_\Lambda}^{1/2},
\ea 
where the cosmic densities $\Omega_i$ have their usual meaning. In this work, we will assume that $f\sub{sky}=1$. 

%

The lower bound in the $\D\log M$ integration in Eq. \re{Nto} is the minimum PBH mass (at formation time) below which a PBH would have evaporated by redshift $z$. For $z=0$, it is well known that 
\ba M\sub{min}(z=0)=5.1\times10^{14}\text{ g}\approx 2.6\times10^{-19} M_\odot,\ea
\citep{hawking}. At higher redshifts, the minimum initial mass can be estimated assuming some basic properties of black holes. We outline the calculations in Appendix A. We found that $M\sub{min}$ remains within the same order of magnitude for a wide range of redshift (see Fig. \ref{fig:mmin} therein). Therefore, for models which generate an observationally interesting abundance of PBHs, it is sufficient to make the approximation $M\sub{min}(z)\approx M\sub{min}(0)$ in Eq. \ref{Nto}. We have checked that this makes no numerical difference for the models studied in this work.








\section{PBH formation: a case study}

\subsection{The log-$\delta$ model}
It is well known that the simplest models of single-field slow-roll inflation cannot produce observable abundance of PBHs unless the primordial power spectrum is very blue (although this has firmly been ruled out by CMB constraints). Viable inflation models which generate an interesting density of PBHs are potentials that typically produce sharp features in the primordial power spectrum, so as to generate power at small scales (see \cite{kawasaki,garcia2,drees, pi, mishra} for some theoretical models). In this work, we will represent a generic primordial power spectrum with a sharp feature using a delta function spike in $\ln k$, \ie
\ba \mc{P_R}(k)= A\delta_D(\ln k - \ln k_0),\ea
where $\delta_D$ is the Dirac delta function. The constants $A$ and $k_0$ parametrize the amplitude and location of the  spike in the resulting  matter power spectrum. This log $\delta$-function model was previously studied in  \cite{wang} in the context of gravitational wave production by PBHs. 

\subsection{Observational constraints}\lab{mono}

A range of observational constraints, including CMB anisotropies \citep{poulter,aloni} and microlensing observations \citep{green,niikura}, have placed upper bounds on $f(M)$, \ie\ the PBH fraction in CDM (see, for example, \cite{carr2, carr3}). Nevertheless, the published bounds assume that all PBHs have the same mass. These so-called \ii{monochromatic} constraints on $f(M)$ were traditionally the main quantity of interest in the literature, as there is a wide range of observational techniques that can place upper bounds on $f(M)$ over several decades of $M$.

If we now assume that PBHs are formed across a spectrum of masses, the monochromatic upper bounds, denoted $f\super{mono}\sub{max}(M)$, must be corrected using procedures such as those previously presented in \cite{carr2, kuhnel, azhar,lehmann}. These studies have only relatively recently gained traction, but are nevertheless indispensable if PBHs were to be taken as a serious candidate for dark matter and GW sources.

\begin{figure} 
   \centering
   \includegraphics[width=0.5\textwidth]{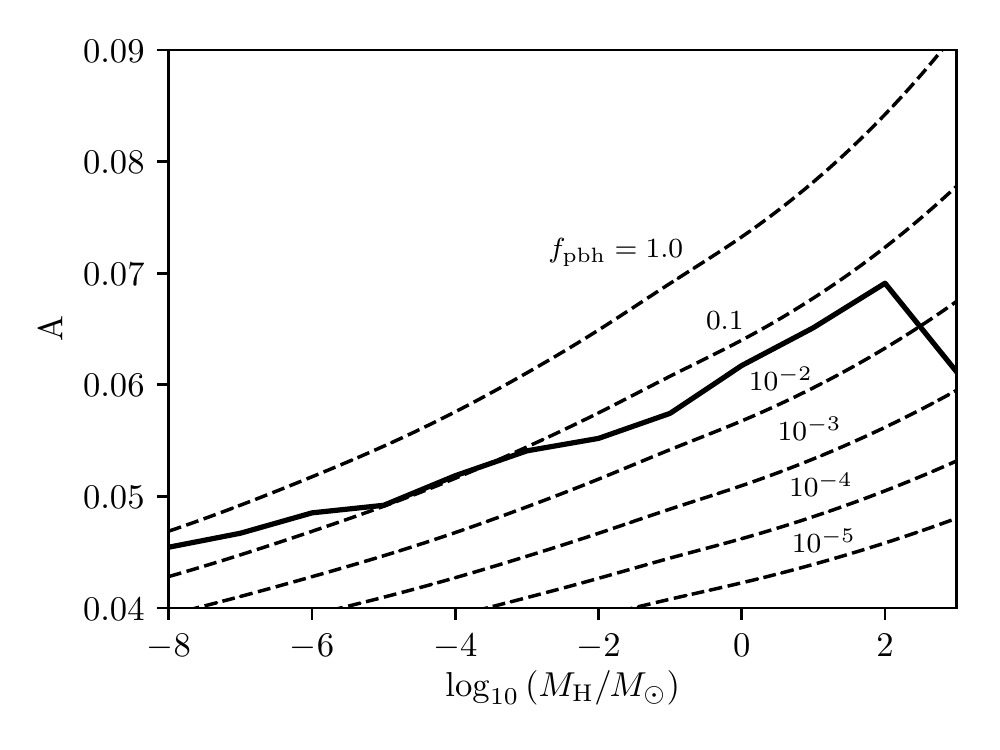} 
   \caption{The dashed lines show the theoretical values of the PBH to CDM ratio, $f\sub{PBH}$ (Eq. \ref{fint}) as a function of the parameters $A$ and $M_H$ in the log-$\delta$ model. The thick line indicates the observational upper bound $f\sub{PBH,max}$ (Eq. \ref{correct}) converted from monochromatic constraints in the literature.}
   \label{fpbh}
\end{figure}

The upshot from these studies is that the corrected upper bound for the total PBH fraction in CDM, $f\sub{PBH,max}$, is given by 
\ba f\sub{PBH,max}= \bkt{\int {f(M)\over f\super{mono}\sub{max}(M)} \D\log M}^{-1},\lab{correct}\ea
\cite{carr2}. The result from applying this correction to monochromatic constraints on the log-$\delta$ model is shown in Fig. \ref{fpbh}. The figure shows the contour lines of constant PBH fraction $f\sub{PBH}$ (Eq. \ref{fint}), as a function of model parameter $A$ (vertical axis) and $k_0$ (horizontal axis, converted to the corresponding horizon mass through Eq. \re{kh}). The thick line shows the corrected upper bound $f\sub{PBH,max}$. In other words, the region below the thick line is the allowed parameter space for the log-$\delta$ model given current observations\footnote{We use observational constraints summarised in Fig 3 of \cite{carr}, and interpolated the upper bounds to obtain an approximate functional form for  $f\super{mono}\sub{max}(M)$.}.

The upper bound is increasing in the domain shown, until $M_H\sim10^2M_\odot$, where the dip corresponds to the more stringent constraint from the CMB anisotropies, since PBH accretion effects can significantly alter the ionization and thermal history of the Universe \citep{ricotti}.

Another interesting observation from the figure is the values of $f\sub{pbh}$ along the thick line. The maximum occurs when the spike is at $M_H=10^{-8}M_\odot$, with $f\sub{PBH}\approx0.46$, and the minimum at $M_H=10^3M_\odot$, with $f\sub{PBH}\approx1.6\times10^{-3}$. This means that present constraints allows the log-$\delta$ model to consolidate almost half of all dark matter into PBHs. However, this comes from imposing a spike at very small scales where the additional nonlinear effects (which have been unaccounted for) become significant. These small-scale effects include large PBH velocity dispersion, accretion and clustering effects seen in previous numerical investigations (\eg \cite{hutsi,inman}). These effects will weaken the validity of the upper bounds on $f\sub{pbh}$ at such small scales.


\section{Extreme PBHs}

Having established a method to calculate the PBH number count and mass function, we now set out to derive the probability distribution of the most massive PBHs expected in an observational volume. Our calculation is based on the exact extreme-value formalism previously used in the context of massive galaxy clusters \citep{harrison,harrison2} and cosmic voids \citep{me_void1, me_void2}.  We summarise the key concepts and equations in this section. 

Using the most massive PBHs to constrain their cosmological origin is motivated by the same reasons that the most massive galaxy clusters and  the largest cosmic voids can be used to constrain cosmology: the largest and most massive structures can typically be observed more easily whilst smaller objects are more dynamic and their observation typically suffers from larger systematic errors. In terms of PBHs, extreme-value probabilities can, at least, constrain the parameters of the underlying inflationary theory, or shed light on their merger history, or, at best, rule out the formation theory altogether. Whilst a complete mass distribution of PBHs within an observational volume would be an even more powerful discriminant of PBH formation theories, in practice it would be extremely challenging to determine with certainty which black holes are \ii{primordial} and which are formed through a stellar collapse or a series of mergers (see \cite{garcia,chen} for some novel methods).

\subsection{Exact extreme-value formalism}

From the PBH number count \re{Nto}, we can construct the probability density function (pdf) for the mass distribution of PBHs with mass in the interval $[\log M,\log M+\D\log M]$ within the redshift range $[0,z]$ as
\ba
f_{<z}(M)= {f\sub{sky}\over N\sub{tot}}  \int^{z}_{0} \textrm{d}z  \diff{V}{z} \diff{n}{\log M}.\lab{pd}
\ea
To verify that this function behaves like a pdf, one can see that by  comparing \re{Nto} and \re{pd}, we have the correct normalization
$$\int_{-\infty}^{\infty} f_{<z}(M)\,\D\log M=1.$$

The cumulative probability distribution (cdf), $F(M)$, can then be constructed by integrating the pdf as usual:
\ba F(M)=\int_{\log M\sub{min}}^{\log M} f_{<z}(m)\, \D \log m.\ea
This gives the probability that an observed PBH has mass $\leq M$.

Now consider $N$ observations of PBHs drawn from a probability  distribution with cdf $F(M)$. We can ask: what is the probability that the observed PBH will all have mass $\leq M^*$? The required probability, $\Phi$, is simply the product of the cdfs:
\ba \Phi(M^*,N)=\prod_{i=1}^N F_i(M\leq M^*)=F^N(M^*),\ea
assuming that PBH masses are independent, identically distributed variables. As $\Phi$ is another cdf, the pdf of \ii{extreme-mass} PBH can be obtained by differentiation:
\ba \phi(M^*,N)=\diff{}{\log M^*}F^N(M^*)=Nf_{<z}(M^*)[F(M^*)]^{N-1}.\lab{eev}\ea
It is also useful to note that the peak of the extreme-value pdf (the turning point of $\phi$) is attained at the zero of the function
\ba X(M) = (N-1)f_{<z}^2 +F \diff{f_{<z}}{\log M},\lab{XX}\ea
as can be seen by setting $\text{d}\phi/\text{d}\log M^*=0$. 

In summary, starting with the PBH mass function, one can derive the extreme-value pdf for PBHs using Eq. \ref{eev}.

\subsection{Application to the log-$\delta$ model}

\begin{figure} 
   \centering
   \includegraphics[width=0.5\textwidth]{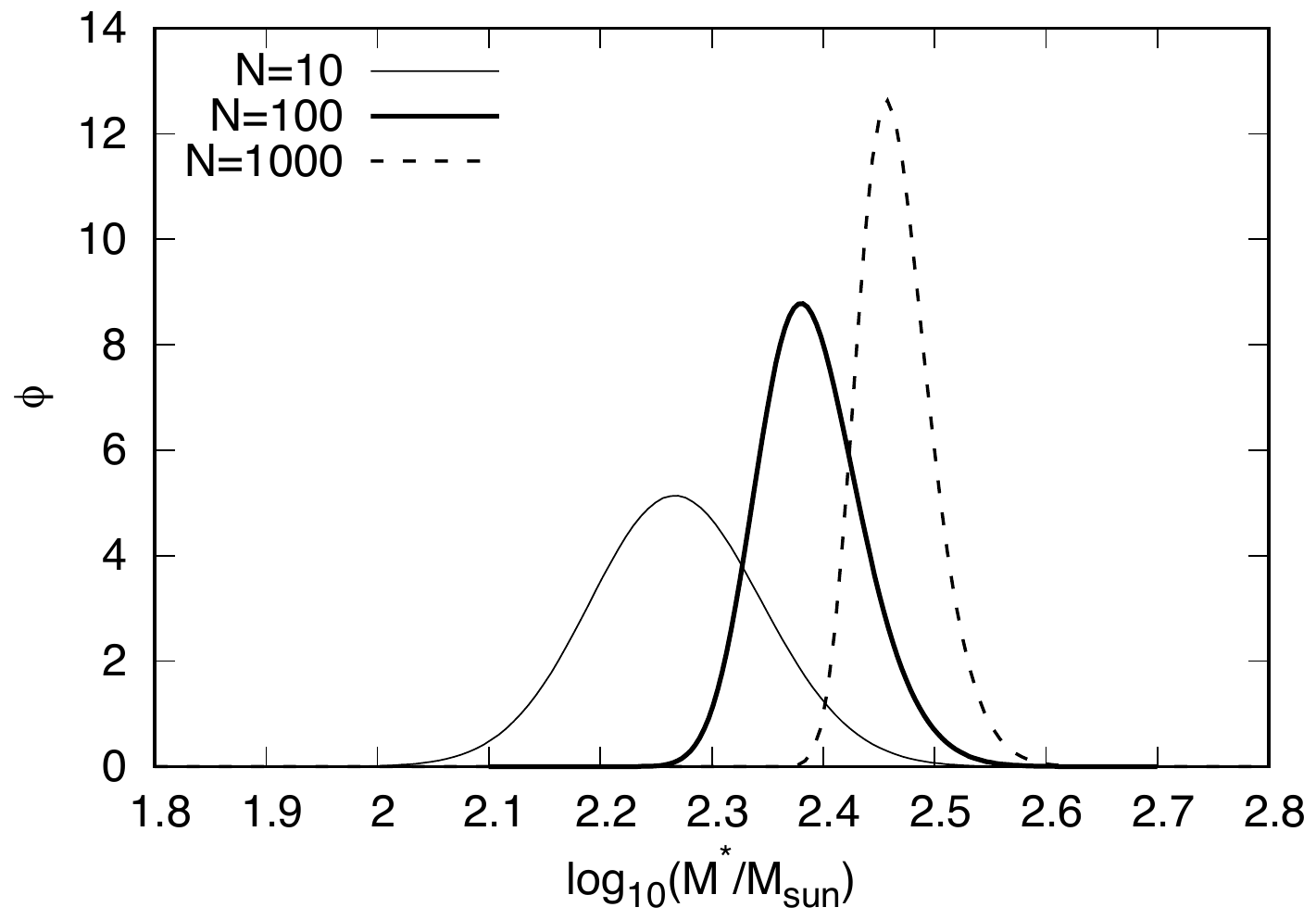} 
   \caption{The extreme-value probability density function for PBHs assuming the log-$\delta$ model with spike at $M_H=10^2 M_\odot$, assuming that $N=10$, $10^2$, $10^3$ observations up to $z=0.2$. }
   \label{extpbh}
\end{figure}

Figure \ref{extpbh} shows the pdfs of extreme-mass PBHs given for $N=10^2 ,10^3$ and $10^4$ observations up $z=0.2$ (this figure summarises the key results of this work). We assume the log-$\delta$ model with the power-spectrum spike at $M_H=10^2M_\odot$. The pdfs are not symmetric but have a positive skewness, consistent with previous derivations of extreme-value pdfs \cite{me_void1, me_void2}. As $N$ increases, the peaks of the pdf naturally shifts towards higher values of $M^*$, with increasing kurtosis (\ie\ more sharply peaked).

\begin{figure} 
   \centering
   \includegraphics[width=0.5\textwidth]{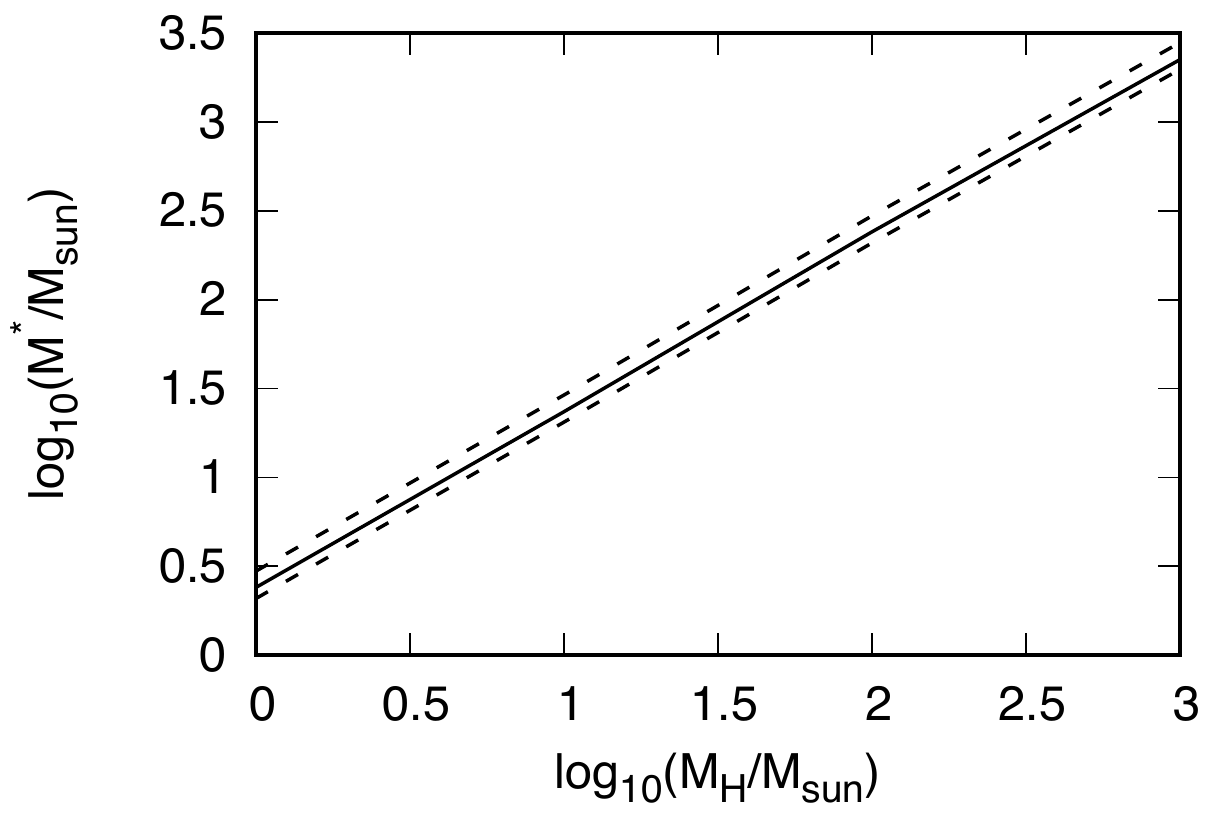} 
   \caption{Profile of the extreme-value pdf peaks (solid line) and the 5th/95th percentiles (dashed lines) for the log-$\delta$ model as the spike location is varied (horizontal axis), assuming 100 observations of PBHs up to $z=0.2$. The vertical axis shows the location of the peaks. The relationship is linear to a good approximation (Eq. \ref{lin}).}
   \label{band}
\end{figure}

When we vary the location of the spike (whilst keeping $N$ fixed, and using values of $A$ that saturate the upper bound shown in Fig. \ref{fpbh}), we obtain an almost linear variation as shown in Fig. \ref{band} (in which $N=100$). Each vertical slice of this figure can be regarded as the profile of the extreme-value pdf, with the peak of the pdf being along the solid line, whilst the 5th and 95th percentiles are shown in  dashed lines. The band is linear to a good approximation, with the peak $M^*\sub{peak}$ satisfying the relation
\be M^*\sub{peak}\approx 2.3M_H.\lab{lin}\ee
The percentile band spans a narrow range of logarithmic masses. We see that the log-$\delta$ model can produce massive PBHs with masses of order $\sim30M_\odot$, using spikes at $M_H\sim10M_\odot$ (the former being within the 5th and 95th percentile band). It is possible to integrate the extreme-value pdfs in figure \ref{extpbh} to calculate the probability that the extreme-mass PBH at redshift 0.2 is, say, $>30M_\odot$.

\begin{figure} 
   \centering
   \includegraphics[width=0.5\textwidth]{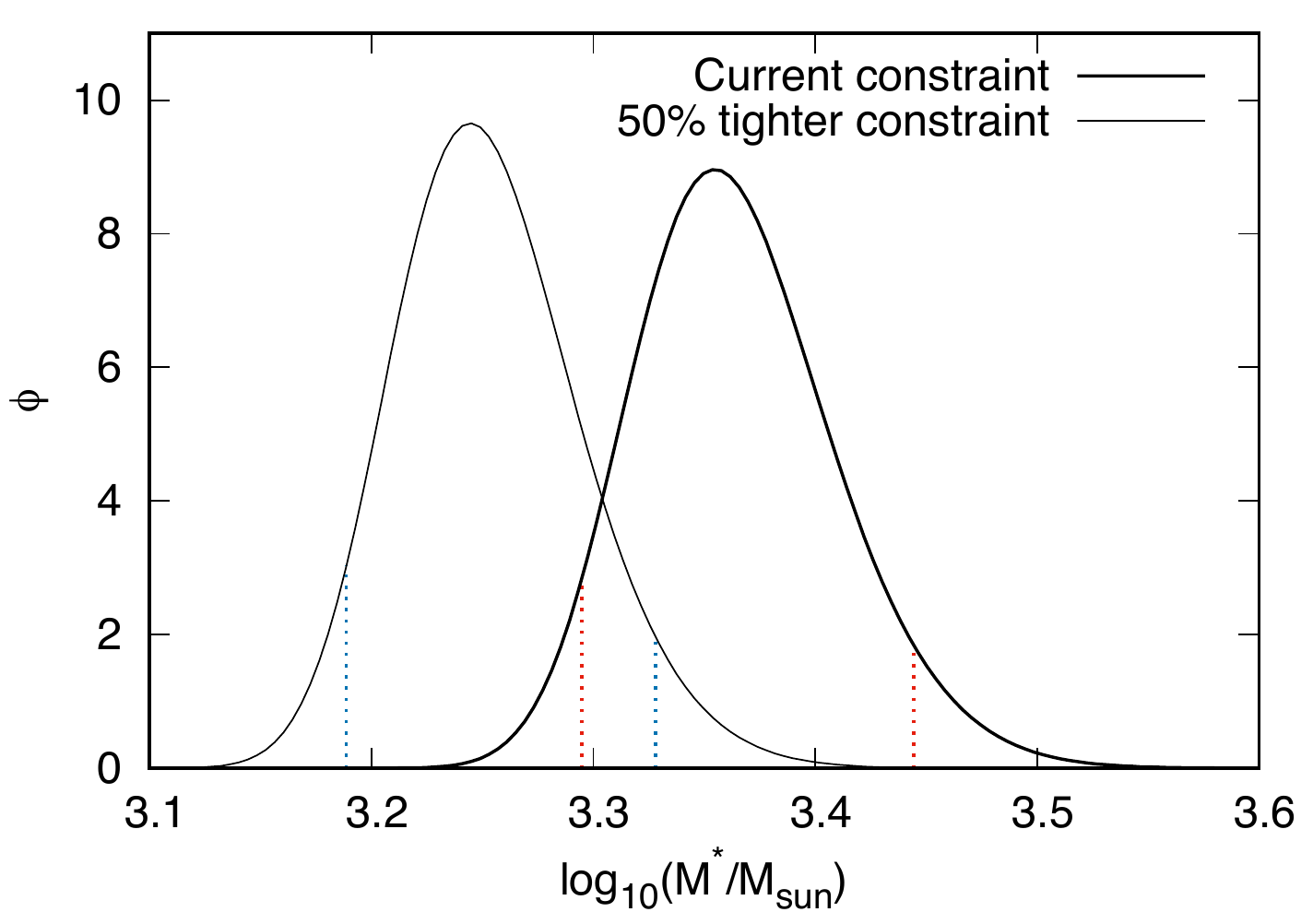} 
   \caption{Extreme-value pdfs for the log-$\delta$ model with $M_H=10^3M_\odot$ given current constraints (thick line) and futuristic  constraints. The pair of vertical dotted lines on each pdf indicate the 5th and 95th percentiles. If observational constraints on $f\sub{pbh}$ were tightened to 50\% the current values, the peak of the extreme-value pdf would shift downwards by $20\%$, whilst the inter-percentile distance would shrink by $\sim30\%$. }
   \label{shift}
\end{figure}

It is also interesting to consider how tightening observation bounds will affect the extreme-value pdfs. Fig. \ref{shift} shows what happens in this situation in the model with $M_H=10^3M_\odot$ (with $N=100$), supposing that the upper bound on $f\sub{pbh}$ is tightened to 50\% of the current values (a realistic prospects for future experiments such as Euclid \citep{habouzit}). We see that, in line with expectation, the pdf shifts to smaller masses by $\sim20\%$, whilst the distance between the 5\% and 95\% percentiles shrinks by $\sim30\%$. 

Finally, one might ask what value of $N$ should be used in this kind of study.  Although from a statistical point of view, $N$ is defined as the number of distinct samples drawn the pdf $f(M)$, in practice it is unclear how to quantify the true number of PBH observations and especially given the additional complication of various selection biases. Individually identified PBHs can be detected using different probes such as microlensing and gravitational-wave emission, but the observable number of PBHs detectable  by one particular method is much smaller than the total number that would be theoretically observable. Thus, to minimise the effect of selection bias, it is more precise to define $N$ as the observed observable number of PBHs from a particular choice of observation. 

\section{Discussion on recent theoretical development}

\subsection{Peak Theory}\lab{peakt}

To complete our investigation, we consider an alternative to calculating PBH abundances using \ii{Peak Theory} which postulates that PBHs result from peaks in the primordial overdensity field exceeding a threshold value (see \cite{bbks,green2} for reviews). It is well known that the PS and Peak Theory do not agree, although previous authors have  suggested that Peak Theory is grounded on a firmer theoretical footing, and is more sensitive to the shape of the inflationary power spectrum \citep{germani, young3, kalaja}. Nevertheless, there are still conceptual issues with both PS and Peak Theory, with a number of extensions having been recently proposed \citep{suyama, germani2}.

In this section, we recalculate the extreme-value distribution $\phi(M)$ shown previously in Figure \ref{extpbh} using Peak Theory in the formulation proposed by Young, Musco and Byrnes \cite{young3}. Using their formalism, we found the PBH fraction $f(M)$ (see Eq. \ref{massfunc}) for the log-$\delta$ model to be\footnote{We note that the power spectrum of the log-$\delta$ model significantly reduces the complicated integrals obtained in \cite{young3}. For example, in their, we find a simple relation $\mu=k_0\sigma$.}
\ba
f\sub{peak}(M)&={\Omega_m\over\Omega\sub{CDM}}\int^{\infty}_{a(M)}\tau(M_H)B\super{peak}_{M_H}(M) \D\log M_H\lab{ppk}\\
B\super{peak}_{M_H}(M)=&= {M\over3\pi M_H}\bkt{k_0\over aH}^3\nu^3 e^{-\nu^2/2},\\
\nu&={\delta(M)\over\sigma(M_H)} ={\bkt{M\over K M_H}^{1/\gamma}+\delta_c\over \sigma(M_H)}\\
a(M)&=\log\bkt{M\over K({2\over3}-\delta_c)^\gamma},\ea
where $\tau(M_H)$ is given in Eq. \re{thermo}. 

\begin{figure} 
   \centering
   \includegraphics[width=0.5\textwidth]{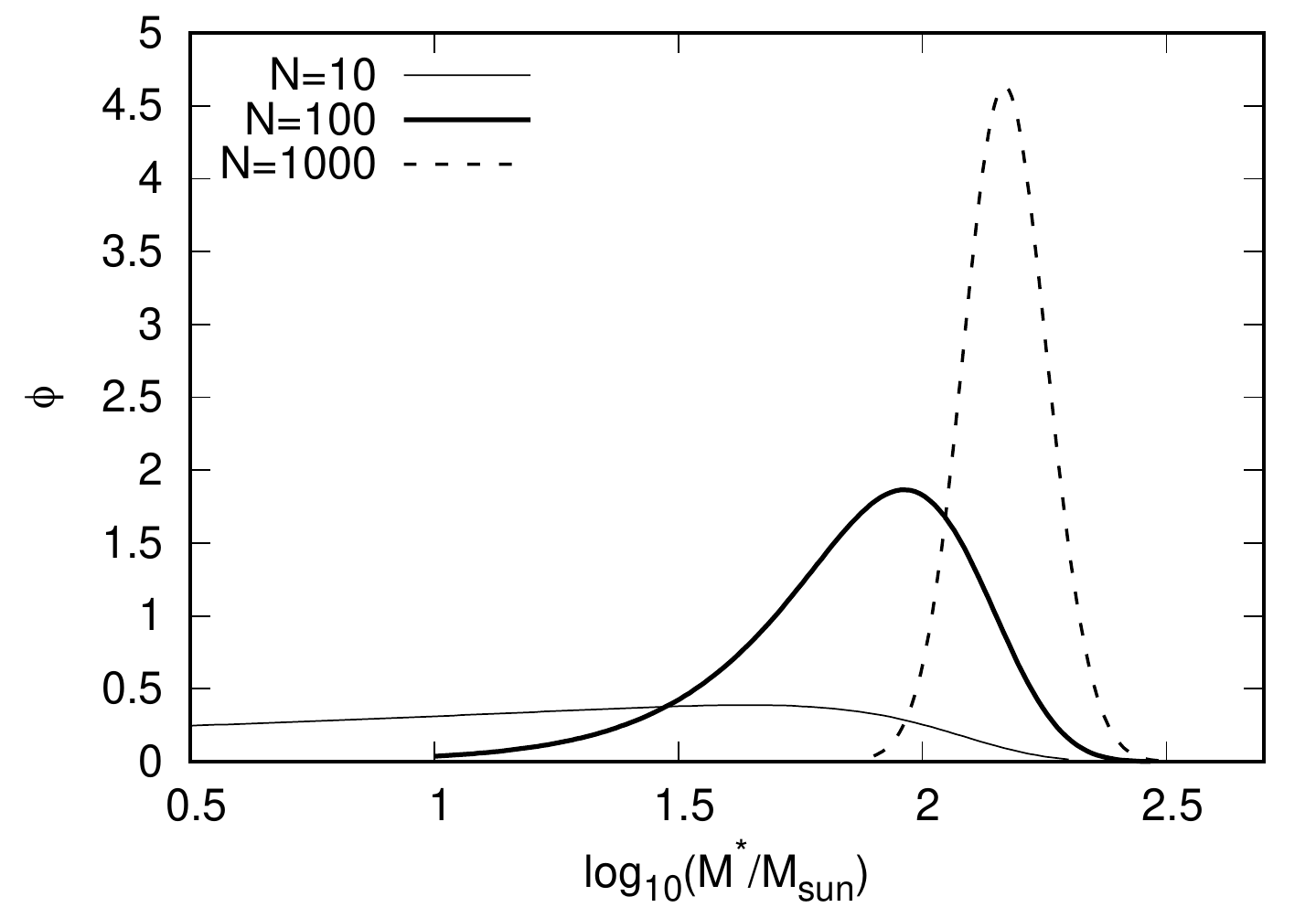} 
   \caption{The extreme-value probability density function from Peak Theory, assuming the log-$\delta$ model with spike at $M_H=10^2 M_\odot$, assuming that $N=10$, $10^2$, $10^3$ observations up to $z=0.2$. In contrast with the pdfs in Fig. \ref{extpbh} (obtained using PS theory), the Peak-Theory pdfs attain maxima at lower $M^*$, and have skewness of the opposite sign.}
   \label{extpbhpeak}
\end{figure}

Some extreme-value pdfs from Peak Theory are shown in 
Fig. \ref{extpbhpeak}. This should be compared with the same distributions calculated using PS formalism in Fig. \ref{extpbh}. 

With $N\gtrsim100$ observations, the extreme-value pdfs from both formalisms attain similar profiles. We observed that the Peak Theory pdfs attain maxima at a slightly lower $M^*$ values compared to the Press-Schechter pdfs. With $N=100$, a similar numerical analysis of the relation between the maxima of the extreme-value-pdf ($M^*\sub{peak-PT)}$) and the location of the spike is found to be:
\be M^*\sub{peak-PT}\approx M_H.\lab{linpeak}\ee
(compare with Eq. \ref{lin}). Although we did observe that the two formalisms predict total PBH number counts that are different by a few orders of magnitude (as corroborated by previous studies), the extreme-value pdfs are not so drastically different. This is because the extreme pdfs are integrated over redshifts and masses, thus when the pdfs are normalised, the effects from large differences in $N\sub{tot}$ are suppressed.

It is also interesting to note that the different functional forms of $B_{M_H}$ lead to different skewness in the extreme-value pdfs (where the skewness is calculated on semilog scale as shown). The Peak-Theory pdfs are negatively skewed whilst the PS pdfs are positively skewed.  See the Appendix B  for an analytic explanation. It may be possible to use this property to distinguish between the peak-theory and Press-Schechter-like formalism of PBH formation.

\subsection{The overdensity profile}\lab{profile}

\begin{figure} 
   \centering
   \includegraphics[width=0.5\textwidth]{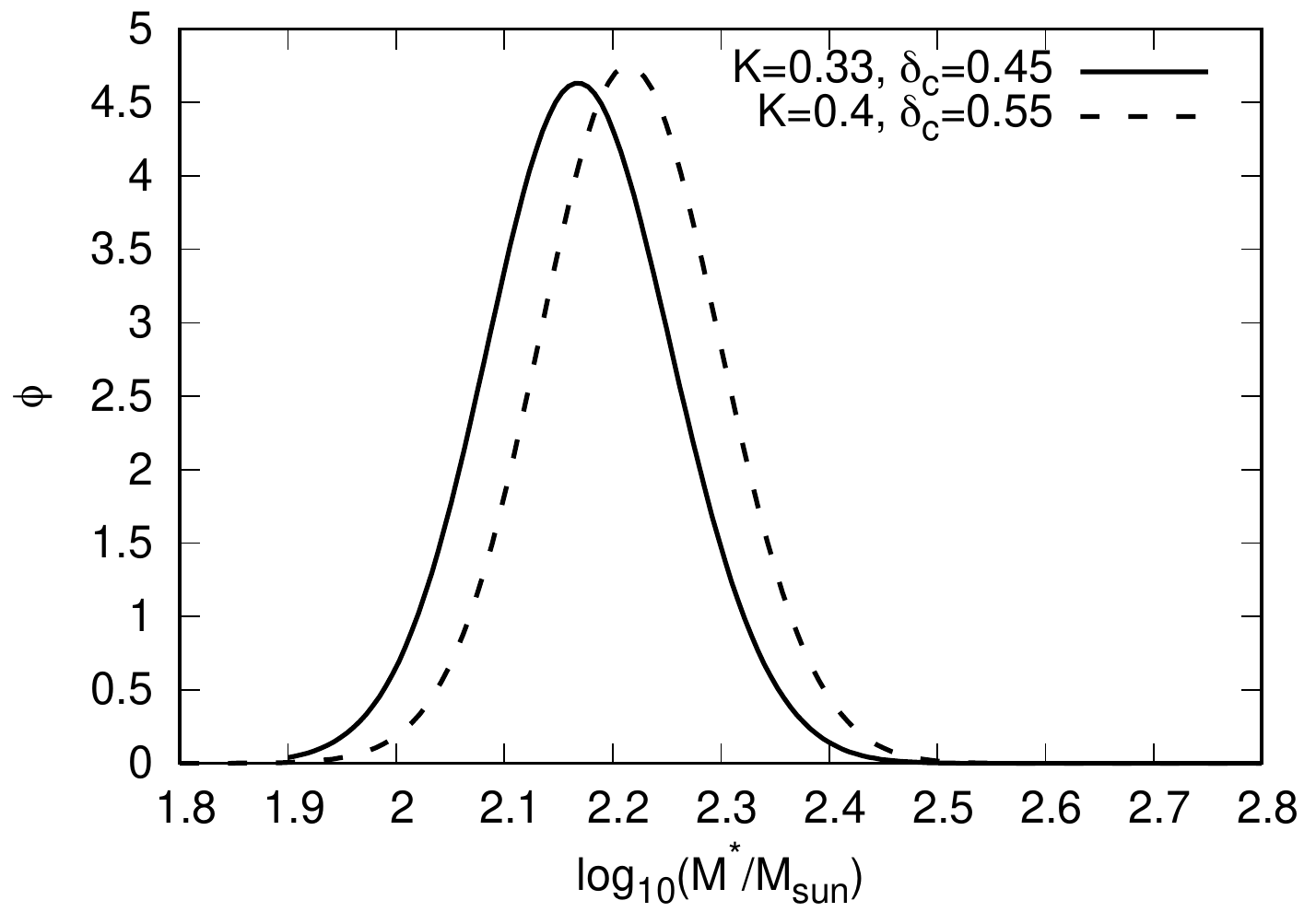} 
   \caption{The effect of changing $K$ and $\delta_c$ on the extreme-value probability density function from Peak Theory, assuming the log-$\delta$ model with spike at $M_H=10^2 M_\odot$, assuming that $N=10^3$ observations up to $z=0.2$. Changing $(K,\delta_c)$ from $(3.3, 0.45)$ to $(4, 0.55)$ (solid and dashed lines respectively) shifts the pdf to higher $M^*$ by a few percent.}
   \label{figalt}
\end{figure}

The shape and height of the profile of density peaks are governed by the constant $K$ and the critical density $\delta_c$ in equation \ref{fraction}. Both quantities can vary depending on typical profiles of the density perturbation (see \cite{musco2} for a comprehensive theoretical study). To this end, we re-evaluate the Peak Theory pdfs using the values $K=4$ and $\delta_c = 0.55$ as proposed by \cite{young3}, instead of the fiducial values $K=3.3$ and $\delta_c=0.45$. Figure \ref{figalt} shows the comparison between the two sets of parameters for the model with $M_H=10^2M_\odot$ assuming $N=10^3$. It appears that the extreme-value pdfs only depend weakly to changes in $K$ and $
\delta_c$: Increasing these parameters (by $\sim30\%$) only results in a few-percent shift of the pdf to higher logarithmic masses.

\section{Conclusion}

In this work, we have established a framework to calculate the  mass distribution of most massive PBHs expected within a given observational volume. The calculations were based mainly on four main ingredients: 
\bit
\item the PBH formation mechanism (\eg details of inflation or the shape of $\mc{P_R}(k)$)
\item the abundance of massive objects (\eg Press-Schechter or Peak Theory)
\item the exact extreme-value formalism
\item the observational constraints on $f\sub{pbh}$ (the PBH fraction in CDM).
\eit

We applied our formalism to the log-$\delta$ model, a prototype of models with a spike in the power spectrum. Such spikes are generically associated with inflationary models that produce interesting densities of PBH (\eg via a phase transition in the early Universe). Our main results are the extreme-value pdf shown in Fig. \ref{extpbh} and \ref{extpbhpeak}. The fact that the location of the power-spectrum spike is close to the peak of the resulting extreme-value pdf gives assurance that our calculations are sound, and can thus be applied to many inflationary models known to produce PBHs. In future work,  we will present a survey of extreme-value pdfs for a range of inflationary scenarios.

Some avenues for further investigation include studying the effect of changing the mass function (for example, extending the Sheth-Tormen mass function to PBHs \citep{me_pbh}), as well as understanding the role of PBH clustering and merger \citep{kohri, young2,tada, raidal,raidal2} which will serve to strengthen the validity of the extreme-value formalism presented here. We envisage that there are other uses for the extreme-value pdfs that the formalism presented can be adapted, for instance, to quantify distribution of the most massive intermediate-mass black holes that could subsequently seed supermassive black holes at galactic centres \citep{dolgov}. 







\subsection{Acknowledgements} 
We thank the referee for helpful comments, and Sai Wang for his help in the early stages of this paper. We acknowledge further helpful comments and feedback from Joe Silk, Kazunori Kohri, Ying-li Zhang, Cristiano Germani and Hardi Veermae. Many of our numerical computations were performed on the University of Hull's VIPER supercomputer.

\appendix

\section*{Appendix A: The minimum initial mass of an unevaporated black hole at redshift $z$.}

Consider a Schwarzschild black hole. Its decay rate depends on three variables, namely, 1) the spin ($s$) of the particles it decays into, 2) the energy ($E$) of those particles, and 3) the instantaneous mass $(M)$ of the black hole. By summing over all the emitted particles, the decay rate of a black hole can be expressed as
\begin{equation}
	\label{pintegral}
	\frac{{\rm d}M}{{\rm d}t} = -\frac{1}{2\pi \hslash c^2} \sum_{j} \Gamma_j \int {\rm d}E \frac{E}{\exp\big(8\pi G E M/\hslash c^3 \big) - (-1)^{2s_j}},\tag{A1}
\end{equation}
where the sum is taken over all emitted particle species. The integral is taken over $(0,\infty)$ for massless particles, or $(\mu_j,\infty)$ for massive particles with rest energy $\mu_j$. $\Gamma_j$ is the dimensionless absorption probability, and  $s_j$ is the spin of the  $j$th species.

\begin{figure}[h] 
   \centering
   \includegraphics[width=0.5\textwidth]{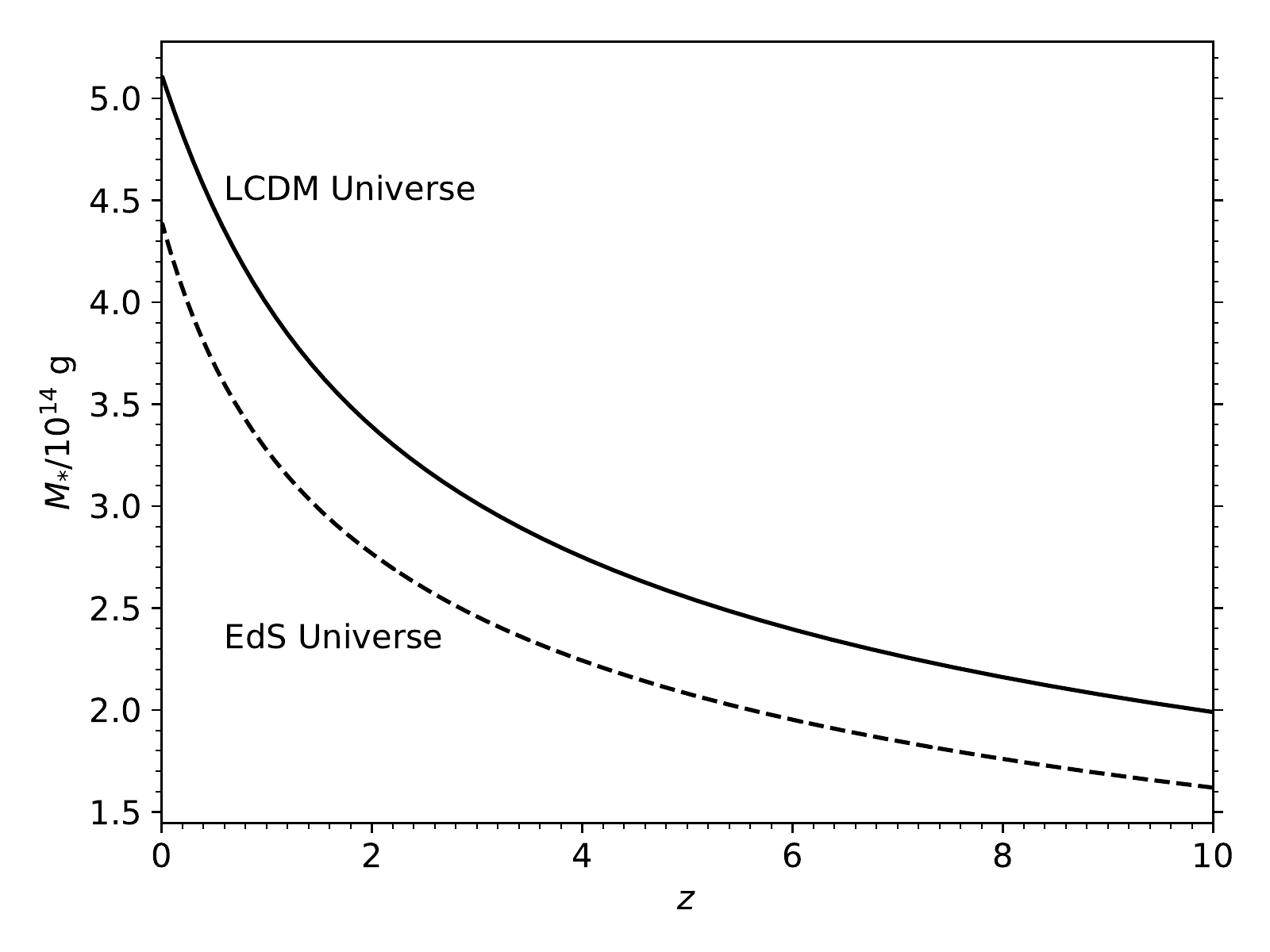} 
   \caption{\label{fig:mmin} Minimum initial mass of a black hole which have not evaporated, observed at redshift $z$.  The solid line is the result assuming the LCDM universe with Planck 2018 parameters. For comparison, the dashed line assumes the Einstein-de Sitter (EdS) cosmology $(\Omega_m=1, \Omega_\Lambda=0)$.}
   \label{Mmin}
\end{figure}

MacGibbon \citep{MacGibbon1991} showed that Eq.~(\ref{pintegral}) can be written as
\begin{equation}
	\label{macgibbon1}
	\frac{{\rm d}M}{{\rm d}t} = -5.34\times10^{22} f\sub{emit}(M) M^{-2} \text{ kg s}^{-1}.\tag{A2}
\end{equation}
The function $f\sub{emit}(M)$ is given in a rather complicated piecewise form in Eq.~(7) in \cite{MacGibbon1991}.   Eq.~(\ref{macgibbon1}) can be inverted and integrated to yield the evaporation timescale, $\tau_{\rm evap}$, as
\begin{equation}
	\tau_{\rm evap} = \big(1.87266\times10^{-23}\ \text{s kg}^{-1} \big) \int_{M_f}^{M_i} {\rm d}M f(M)^{-1} M^2,\tag{A3}
\end{equation}
where $M_i$ is the initial mass of the black hole and $M_{f}$ is the final mass. Letting $M_f = 0$ and $M_i = M_*$, we can obtain $M_*$ as a function of $z$ by solving the non-linear equation:
\begin{equation}
	\tau_{\rm evap}(z)\bigg|_{M_f = 0, M_i = M_*} = t_{\rm univ}(z),\tag{A4}
\end{equation}
where $t_{\rm univ}$ is the age of the universe at redshift $z$.  The minimum initial black hole mass as a function of redshift is shown in Figure~\ref{fig:mmin}. The figure closely resembles Fig. 1 of \cite{MacGibbon1991}, although we believe that the labelling of the two  curves in that figure should be exchanged.

\section*{Appendix B: Skewness of the extreme-value distribution (Press-Schechter vs. Peak-Theory).}

Let us explore a simple analytic approximation of the extreme-value pdf in order to show that the skewnesses (on log mass  scale) calculated using the above theories are of opposite signs. To this end, we will study the simplest case $N=1$, in which case the extreme-value pdf is $\phi(M)=f_{<z}(M)\sim f(M)/M$ (using Eqs. \ref{massfunc}, \ref{pd} and \ref{eev}). Here $f(M)$ are the PBH fractions given in \ref{fM} and \ref{ppk} for the two formalisms respectively.  We will focus on the evolution in $M$ only.

\bb{Press-Schechter}: We make the approximation for the variance of PBH-forming overdensity: $\sigma\sim R^{-2})$ where $R$ is the scale of the window function. Thus, in terms of mass within the filter, $\sigma\sim M^{-2/3}$. The mass-dependent terms in the mass fraction \ref{fM} can then be expressed as  
\[f\sub{PS}(M)\sim M^4\int x^{-13/ 3} \exp\bkt{-x^{4/3}{\bkts{\delta_c+ {M^3\over x^3}}}^2}\D x.\]

We approximate the integral in two regimes. When $M$ is small, the $M$ dependence in the integral is negligible, so $f\sub{PS}\sim M^4$. When $M$ is large, the $M$ dependence can be removed from the integral via a substitution $t=M^6x^{-{14/3}},$ leaving a gamma function and an overall $M$ dependence of  $f\sub{PS}\sim M^{-2/7}$. In summary, we find
\[\phi\sub{PS}(M)\sim \begin{cases}
M^3& , M\ll1,\\
M^{-9/7}&, M\gg1.
\end{cases}\]

\bb{Peak-Theory}: A similar approximation scheme [but this time retaining the integral due to the $M$ dependence in the lower limit of \re{ppk}] shows that the mass fraction can be approximated by an incomplete gamma function. One then obtains:
\[\phi\sub{peak}(M)\sim M^{3/2}\,\Gamma\bkt{{25\over8}, M^{4/3}}.\]

The functions $\phi\sub{PS}$ and $\phi\sub{peak}$ are plotted in figure \re{gamma} below on semilog scale. Evidently their skewnesses differs in sign, as can be confirmed numerically. For larger $N$, the extreme-value pdfs retain their respective sign as seen in the main text.

\begin{figure}[h] 
   \centering
   \includegraphics[width=0.4\textwidth]{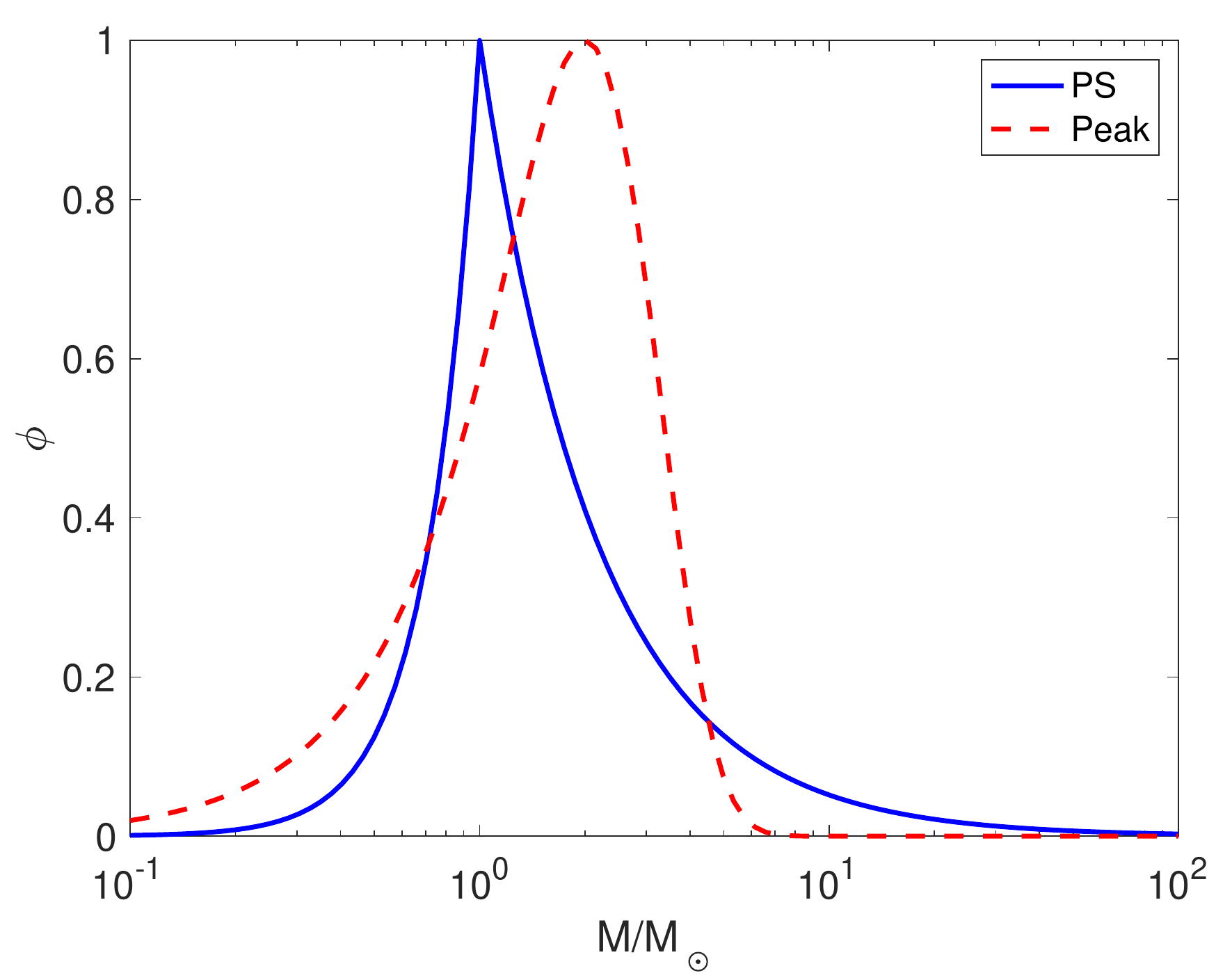} 
   \caption{\label{gamma} The simplified extreme-value pdf for the Press-Schechter and Peak-Theory formalisms, normalised so that the maximum is at $\phi=1$. Their skewnesses (on log mass scale) are evidently of opposite signs.}
\end{figure}

\bibliography{pbh_ref}

\end{document}